\documentclass[aps,prb,reprint,showpacs,groupedaddress,amsfonts,floatfix]{revtex4-2}
\usepackage{CJK}
\usepackage{graphics}

\usepackage[dvips]{graphicx}
\usepackage{amsmath}

\begin{document}
\begin{CJK*}{GB}{} % Use default fonts from CJK (see below)
%
%   TITLE
%

\title{Stopping power of electrons in a semiconductor channel for swift point charges}
%: \\
%Scattering modeling based on regularized interaction}

%
%  AUTHORS AND AFFILIATIONS
%

%
\author{I. Nagy}
\affiliation{Department of Theoretical Physics,
Institute of Physics, \\
Budapest University of Technology and Economics, \\ H-1521 Budapest, Hungary}
\affiliation{Donostia International Physics Center, P. Manuel de
Lardizabal 4, \\ E-20018 San Sebasti\'an, Spain}
\author{ I. Aldazabal}
\affiliation{Centro de F\'{i}sica de Materiales (CSIC-UPV/EHU)-MPC,
P. Manuel de Lardizabal 5, \\ E-20018 San Sebasti\'an, Spain}
\affiliation{Donostia International Physics Center, P. Manuel de
Lardizabal 4, \\ E-20018 San Sebasti\'an, Spain}

\date{\today}
\begin{abstract}

The nonperturbative kinetic framework for the stopping power of a charged-particle system for swift
point projectiles is implemented. The pair-interaction potential energy required in this framework 
to two-body elastic scattering is based on the screened interaction energy between system particles.
In such an energetically optimized modeling the swift bare projectile interacts with independent screened constituents of a fixed-density interacting many-body target. 
The first-order Born momentum-transfer (transport) cross section is calculated and thus a
comparison with stopping data obtained [Phys. Rev. B {\bf 26}, 2335 (1982)]  by swift ions, $Z_1\in{[9,17]}$ and $(v/v_0)\simeq{11}$, under channeling condition in Si is made.
A quantitative  agreement between the elastic scattering-based theoretical stopping and the experimentally observed reduced magnitude is found. Conventionally, such a reduced magnitude 
for the observable is interpreted, applying an equipartition rule, as inelastic energy loss mediated by a collective classical plasma-mode without momentum transfer to the valence-part. Beyond the leading, i.e., first-order Born-Bethe term ($Z_1^2$), the Barkas ($Z_1^3$) and Bloch ($Z_1^4$) terms are discussed, following the arguments of Lindhard for screened interaction. An extension to the case
of stopping of warm dense plasma for swift charges is outlined as well.

\end{abstract}


\maketitle
\end{CJK*}

\section{The status of the problem}

In this short theoretical note we start with dedicated experiment \cite{Golovchenko82} 
for the stopping of swift ions,  $Z_1\in{[9,17]}$ and $(v/v_0)\simeq{11}$, at channeling conditions 
in Si target.  It was found that the measured data are only about {\it half}
of  the result based on the conventional (linearized self-consistent field) dielectric \cite{Lindhard54} result for the stopping power in an electron gas
\begin{equation}
\frac{dE}{dx}\, =\, { \frac{4\pi n_0\, Z_1^2\, e^4}{m v^2}\, \ln\frac{2mv^2}{\hbar \Omega_{pl}}},
\end{equation}
where $\Omega_{pl}^2\equiv{4\pi n_0 e^2/m}$ is the classical 
\cite{Tonks29} plasma frequency in terms of the number-density $n_0$, electron charge $e$, 
and mass $m$. This classical (independent of $\hbar$) quantity represent a collective 
mode because the restoring force on the displaced charged particles arises from the self-consistent classical electric field generated by the local excess charges \cite{Fetter71}. 

In the consideration of a charge-compensated degenerate electron gas, the prototype model
of free-electron metals, the classical collective mode appears in the equation of motion for the 
operator of charge density fluctuation 
$n_{\bf q}=\sum_{j} \exp(i{\bf q}\cdot{{\bf r}_j})$ at small wavelenght
\begin{equation}
\ddot{n}_{\bf q}\, =\,  -\Omega^2_{pl}\, n_{\bf q}  \nonumber
\end{equation}
The underlying field-theoretic (second-quantized) method rests on the Heisenberg equation of
motion for $\dot{n}_{\bf q}$ and application of the random-phase approximation introduced
in the pioneering normal-mode treatment of an interacting electron gas 
\cite{Pines61,Feynman72}.
According to that pioneering treatment \cite{Pines61}, the number of collective modes ($n'$)
appears  as subsidiary condition on which the ground-state energy of the system depends. The variational 
constraint on that energy result in $(n'/n_0)<<1$ in terms of the number density $n_0$ of electrons. 

For a recent comprehensive discussion of Eq.(1) in plasma physics, we refer to \cite{Brown05}.
Besides, in an earlier detailed Lecture Notes \cite{Bonderup81} which includes the classical 
(i.e., the induced electric field) interpretation of the retarding force, an equipartition in the total stopping power was emphasized. Namely, in a single-pole approach for the response function of the model of a charged electron-gas-at-rest (bosonic) system, there are equal contributions to 
the stopping power, $dE/dx$, from "free" and "resonant" collisions.

In the self-consitent-field approach \cite{Bonderup81}, the system Fourier variables for energy- and 
momentum-change to inelastic processes ($\hbar \omega$ and ${\bf q}$) are reinterpreted
($\hbar \omega={\bf q}\cdot{\bf v}$ ) in terms of an externally fixed variable (${\bf v}$) of a heavy ion. 
On the role of the (undamped) classical collective mode in an {\it inelastic} process, like the external-electron--charged-medium interaction, we refer to a particularly clear discussion in \cite{Fetter71}. Briefly, in such an inelastic process  the momentum transfer and the (positive) energy loss may be 
varied independently. An isolated resonant peak in the dynamical structure function of the system have been observed in the transmission \cite{Pines66} of electrons through certain thin metallic films.

The above-mentioned equipartition in Eq. (1) is based \cite{Bonderup81} on re-writing it into the form 
\begin{eqnarray}
\frac{dE}{dx}\, =&&\, { \frac{4\pi n_0\, Z_1^2\, e^4}{m v^2}\, \ln\frac{q_{max}}{q_{min}}}\nonumber \\
=&&\frac{4\pi n_0\, Z_1^2\, e^4}{m v^2}\, \left[\ln\frac{q_{max}}{q_0}+\ln\frac{q_0}{q_{min}}\right] \nonumber
\end{eqnarray}
To such a momentum-transfer-based representation of the logarithm-argument in Eq.(1) the  maximum and minimum are determined by the two real
solutions of a Bogoliubov-type \cite{Fetter71} (now, for Coulomb forces) energetic constraint \cite{Bonderup81,Nagy93} within the single-pole approximation
\begin{equation}
\left[\frac{(\hbar q)^2}{2m}\right]^2 + (\hbar \Omega_{pl})^2 =(\hbar q v)^2. \nonumber
\end{equation}
The two solutions result in the following mathematical identification of a $q_0$ value
\begin{equation}
\frac{\hbar^2}{2m}(q_{max}\cdot{q_{min}})=\hbar \Omega_{pl}\equiv{\frac{(\hbar q_0)^2}{2m}}, \nonumber
\end{equation}
independently of $v$. In the large velocity limit one gets
$\ln(q_{max}/q_{min})=2\ln(q_{max}/q_0)$.  Thus, an  equipartition rule in 
stopping is demonstrated in this manner. 
Such a  rule is applied in a textbook \cite{Komarov81}
to experimental (reduced) ion-stopping at channeling condition by supposing the dominating 
role of the collective polarization response and excluding binary events. We notice here, however, 
that one may interpret the logarithm-argument in Eq.(1) as the ratio of energy losses in elastic 
binary $(2mv^2)$ and collective inelastic ($\hbar \Omega_{pl}$) excitations.

The motivating experiment \cite{Golovchenko82} on Si, as we mentioned, yields about half of the value
based on the conventional form. Besides, a detailed analysis of the higher-order (with opposite-sign) terms was given by pointing out their efficient cancellation. This leaves the first-order Born scaling $(\propto{Z_1^2})$ as the proper one at the experimental conditions with positive ions. Notice, 
that an about $(1/v^2)$-scaling for the sign-dependent ($\propto{Z_1^3}$) term was predicted  
in CERN-experiment \cite{Andersen89} using swift protons and antiprotons and a Si detector-target. 

This paper is devoted to the problem outlined above. We will use the kinetic theory for the
stopping power in which the swift heavy intruder interacts with system constituent via elastic scattering transferring momentum and energy in independent binary processes. The applied theory is well-established \cite{Bonderup81,Sigmund82,Arista84,Bonig89}. 
However, its implementation needs a reasonable two-particle interaction energy
in which a large part of physics is encoded \cite{Zwicknagel99}. In particular, this interaction
energy should respect the interaction-time aspect of the binary process and the many-body 
aspect of the charged system without the external swift projectile \cite{Nagy20}.

\section{Modeling, results, and discussion}

In the kinetic theory \cite{Bonderup81,Sigmund82,Arista84,Bonig89} one has for the observable, 
i.e., for the stopping power
\begin{equation}
\frac{dE}{dx}\, =\, n_0\, mv^2\, \sigma_{tr}(k),
\end{equation}
at high ion-velocities. The momentum-transfer ($tr$) cross section is thus calculated 
at relative wave-number $k= mv/\hbar$ in quantum mechanics. For swift heavy ($M>>m$)
projectiles we can neglect an averaged recoil term [$\propto{(m/M)}$] of elastic 
binary collisions \cite{Sigmund82,Arista84}. Notice that, in quantum mechanics, the 
kinematically transparent Eq.(2) contains statistical and quantum mechanical averaging. The former appears as a cumulative factor ($\propto{n_0}$) at high velocities, the latter by calculating the expectation value of the force operator \cite{Bonig89,Zwicknagel99} over the set of 
scattering eigen-states with $l\rightarrow{(l+1)}$ selection in matrix elements.

The quantum-mechanical transport cross section ($\sigma_{tr}$) for elastic electron scattering 
in the center-of-mass system with a spherical interaction potential is defined as follows
\begin{equation}
\sigma_{tr}(k)=2\pi\int_{0}^{\pi}d\theta\, \sin\theta\, (1-\cos\theta)|f(k,\theta)|^2,
\end{equation}
in terms of the (complex) scattering amplitude. Its partial-wave representation 
\begin{equation}
f(k,\theta)=\frac{1}{k}\, \sum_{l=0}^{\infty}\, (2l+1)\,  e^{i\delta_l(k)}\, \sin[\delta_l(k)]\, P_l(\cos\theta),
\end{equation}
in the defining expression to an averaged momentum transfer in Eq.(3) results in
\begin{equation}
\sigma_{tr}(k)=\frac{4\pi}{k^2}\sum_{l=0}^{\infty}(l+1)\sin^2[\delta_l(k)-\delta_{l+1}(k)],
\end{equation}
where $\delta_l(k)$ are Bessel phase shifts determined by the scattering Schr\"odinger equation.

Considering the experimental conditions, $(v/v_0)=11$, we apply the first-order Born (B)
approximation where, instead of solving the radial wave equations, one uses unperturbed plane waves
(eigenfunctions of the momentum operator) for incoming and outgoing states. Since in our case the phase shifts are small, $ [\delta_l(k)-\delta_{l+1}(k)]\propto[(1/k)/(l+1)]$, we can write 
\begin{equation}
\begin{split}
f^{(B)}(k,\theta) & = \frac{1}{k}\, \sum_{l=0}^{\infty}\, (2l+1)\,  \delta^{(B)}_l(k)]\, P_l(\cos\theta) \\
 & = \frac{2m}{\hbar^2}\, \int_{0}^{\infty}V(r)\, \frac{\sin(K r)}{K r}\, r^2\, dr \\
 & = f^{(B)}(K)
\end{split}
\end{equation}
where $K=2k\sin(\theta/2)$, and thus $\sin\theta d\theta=(K/k^2) dK$ 
and $(1-\cos\theta)=K^2/2k^2$ to Eq.(3). The above correspondence is based on the definition of the
first-order Born phase shifts, in terms of $V(r)$ and Bessel functions [$j_l(x)$] of the first kind, 
and application of an  identity
\begin{equation}
\frac{\sin(Kr)}{Kr}\, =\, \sum_{l=0}^{\infty}(2l+1)j_l^2(kr)P_l(\cos\theta). \nonumber
\end{equation}

Now, we arrived at the basic task in our modeling. According to the short discussion in the previous Section and in \cite{Nagy20}, we suppose that the swift bare projectile interact with screened independent constituents of the many-body system via $V(r)=-Z_1v(r)$ in elastic scattering. 
In other words, to justify a Sommerfeld-like (free-particle)  picture we employ a simplified version of the heuristic quasiparticle-description of Landau, which is in fact a canonical approximation procedure.
Thus the residual interaction between system particles (without external intruder) is characterized by $v(r)$. It is a residual interaction when all averaging, correlating, and screening effects in the system have been taken into account \cite{Ziman69}. 

Next, we turn to a physics-based design of the interparticle $v(K)$, and refer to a canonical, 
Bogoliubov-type form \cite{Fetter71} for the square of the normal mode excitation ($E_{ex}$) energy
\begin{equation}
(E_{ex})^2=\left[\frac{(\hbar K)^2}{2m^{\star}}\right]^2 + 2v(K)n_0\frac{(\hbar K)^2}{2m^{\star}}
\end{equation}
which might allow a renormalization (field-theoretic) procedure via the effective mass.
Clearly, in order to keep a Sommerfeld-like free-particle picture we get
$v(K\rightarrow{0})\propto{K^2}$. In the opposite limit, i.e., at large $K$, which corresponds to the short-range in real space, one should recover the well-known Coulomb 
form $v(K\rightarrow{\infty})\propto{1/K^2}$. 

The small $K$ limit for $v(K)$,
as Eq.(6) shows, is related to a Friedel-like \cite{Mahan81} sum of phase shifts since at the 
forward limit ($\theta=0$) one has $K=0$ and $P_l(1)=1$ for the Legendre polynomials. Such a sum determines, via the Fumi theorem \cite{Mahan81}, the kinetic-energy-change in the many-body system. At first-order in interparticle $v(r)$ the variational consistency on the ground state requires 
its minimization. And, in harmony with this energetic constraint, the Friedel sum must
be zero since we do not consider an excess charge in the charged host.

Based on the above physics-details, and restricting ourselves to the simplest (i.e., 
one-parametric) form with mathematical (limit) consistency, we write a bosonic 
\cite{Bonch59,Bergara99} form
\begin{equation}
V(K)\, =\, - \, Z_1\, v(K)\, =\, -Z_1e^2\,  \frac{4\pi\, K^2}{K^4 +4\Lambda^4},
\end{equation}
where,  in the equal-mass case \cite{Ladanyi92}, $\Lambda^4=(2\pi n_0)/a_0$ and $(1/a_0)=me^2/\hbar^2$. Thus we get
\begin{equation}
v(r)=\frac{e^2}{r}\, e^{-\Lambda r}\, \cos(\Lambda r).
\end{equation}
Notice, that a Poisson-equation-based classics would give for an "induced" hole-density
$\Delta n(r)=(\Lambda^2/2\pi)(1/r)e^{-\Lambda r}\sin(\Lambda r)$. This is finite at $r=0$, as expected.

At this point we return to the experimentally motivated \cite{Golovchenko82} basic-problem in stopping power of valence electrons of a semiconductor (Si) channel, for swift ions.
The kinetic formalism for elastic binary scattering, implemented by our potential energy, results in
\begin{equation}
\begin{split}
\frac{dE}{dx} & =n_0\, mv^2\, \sigma_{tr}(k) \\ 
& \simeq{n_0\, \frac{4\pi Z_1^2 e^4}{m v^2}}\, \frac{1}{2}\left(\ln\sqrt{1+\frac{4k^4}{\Lambda^4}}-\frac{1}{2}\right) \\
& \simeq{\frac{4\pi n_0\, Z_1^2\, e^4}{m v^2}\, \frac{1}{2}\, \ln\frac{2mv^2}{\hbar \Omega_{pl}}},
\end{split}
\end{equation}
after a quite straightforward quadrature. Here the Bohr unit $(1/a_0)=me^2/\hbar^2$ to $\Lambda^4$ is re-introduced in order to arrive at the final expression, formally in terms of a quantized collective mode 
of a polarizable extended charged medium. Thus we found an explanation, without uncontrolled mixing, to the experimental fact with a {\it half-value} for the observable. Our work may contribute to earlier efforts performed by using quantum mechanical convolution approximation \cite{Azevedo01}, and  classics-based approximation  \cite{Sigmund01} using Bohr modeling. Of course, a trajectory-based classical attempt, i.e., an attempt without angular-momentum quantization in electron scattering, is not able to reproduce the perturbative Born-Bethe limit. Classics may be quantitative
at large enough $Z_1/(a_0 k)$ values for coupling in stopping. We note at this qualitative statement that the semi-classical WKB approach becomes mathematically quantitative \cite{Flugge71} for Coulomb phases [($\sigma_l(Z_1,l,k)$] 
at about $Z_1/(a_0 k)\geq{2}$. The situation in experiments \cite{Golovchenko82} 
was not very far from that range, but from below.

At the end of this Section, we make an other (independent) control of our formalism. 
Following the lead of Lindhard \cite{Lindhard76}, and a strongly related 
analysis \cite{Quinteros91}, on the charge-sign 
($Z_1^3$) effect in stopping we discuss this effect observed \cite{Andersen89}
in Si detector-target by using charge-conjugated swift ($v>5v_0$) particles, 
protons and antiprotons $Z_1=\pm{1}$.
In particular, we address the origin (classical or quantal) of the effect based on the ingredients of the 
leading-order expression. First we observe, following the lead of \cite{Lindhard76},
that the transport cross section scales as ${k^{-4}}\sim{E^{-2}}$, i.e., inversely as the square 
\cite{Quinteros91}
of the large kinetic energy of the scattering particle. Next, the series expansion of $V(r)$ for small distances results in a shift of this kinetic energy giving for it an effective value 
$E_{eff}\simeq{E(1-Z_1e^2\Lambda/E)}$ both to classical 
and quantum treatments \cite{Lindhard76} . Using the such-modified energy to Eq.(10), we get
\begin{equation}
\begin{split}
\frac{dE}{dx} &
\simeq \frac{4\pi n_0\, Z_1^2\, e^4}{m v^2}\, \left[\frac{1}{2}\, \ln\frac{2mv^2}{\hbar \Omega_{pl}} \right. \\
& + \left. \,  Z_1 \frac{e^2}{mv^2}\, (2 \Lambda)\, \ln\left(\frac{2mv^2}{\hbar \Omega_{pl}}\right)\right] ,
\end{split}
\end{equation}
and the second term will be denoted as $Z_1\, L_1$, according to tradition \cite{Lindhard76,Andersen89,Nagy20}.

It is important to compare Eq.(11) with the result based on the higher-order response function
approach \cite{Esbensen90,Pitarke95} for a polarizable medium to characterize the induced 
(retarding) electric field at the swift-projectile position. There, at the  RPA \cite{Lindhard54} level, one obtains 
\begin{equation}
\begin{split}
\frac{dE}{dx} &
\simeq \frac{4\pi n_0\, Z_1^2\, e^4}{m v^2}\, \left[ \ln\frac{2mv^2}{\hbar \Omega_{pl}} \right. \\
& + \left. \, Z_1 \frac{ e^2}{mv^2} \left(\frac{3\, \pi\, \Omega_{pl}}{2 v}\right)\, \ln\left(\frac{2mv^2}{2.13\hbar \Omega_{pl}}\right)\right],
\end{split}
\end{equation}

Figure 1 contains the experimental data denoted by open 
circles and open triangles with error bars. They are taken from Fig. 4 of \cite{Andersen89}, where the 
two corresponding  thickness for Si detector-target were also indicated.  
The solid curve is based on Eq.(11), the dashed one on Eq.(12).
To implement these theoretical expressions, an $r_s=2 a_0$ for the system Wigner-Seitz radius, 
i.e., density parameter, is employed. At that value  $\Lambda\simeq{0.66/a_0}$ 
in our modeling, thus $(\Lambda)^{-1}<{r_s}$. A mathematical equivalence
of the two forms for the
sign-coefficient would result in $\Lambda(v)=(3\pi/4)(\Omega_{pl}/v)<<{(r_s)^{-1}}$.
One might regard  such a deduced form as conveying information on the 
dynamical response of the medium in a stationary (Poisson-equation-based) self-consistent field 
approximation for swift-ion screening. This adiabatic parametrization in the first
term of Eq.(10) would results in its numerical doubling in contradiction with channeling data. 
And a large volume, $V\propto{[\Lambda(v)]^{-3}}$ with
interacting system particles, is excluded from local binary elastic scattering.
Mathematical tuning of $\Lambda$ results in a re-balancing \cite{Nagy20} of the two leading terms in the perturbative stopping power. 
\begin{figure} 
\includegraphics[width=.45\textwidth]{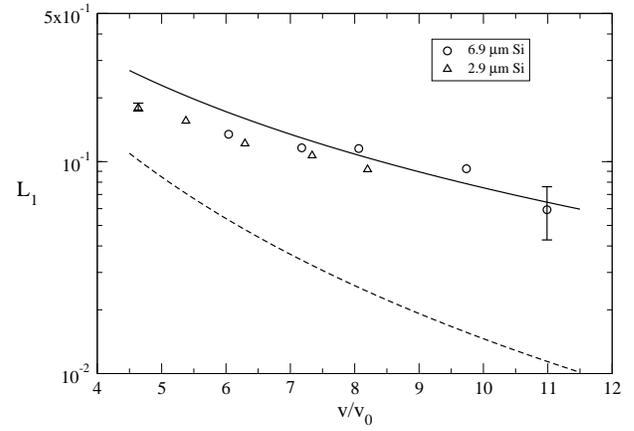}
\caption{Charge-sign coefficients, $L_1$ in Eqs.(11-12), as a function of the swift
projectile (proton and antiproton) velocity. 
The solid curve refers to our result in Eq.(11), the dashed one is based on Eq.(12).
Symbols with error bars refer to data \cite{Andersen89} obtained at CERN for Si detector-target.}
\label{figure1}. 
\end{figure}

Our form for the $L_1$ sign-dependent term is in reasonable harmony with data. Especially,
its experimentally emphasized $(1/v^2)$-dependence is gratifying. The dielectric modeling
gives a quite different, $(1/v)^3$-scaling. It underestimates the data by taking the same density
for the homogeneous host system. The common argument \cite{Esbensen90,Pitarke95}, to cure this underestimation, rests on normalized-weighting over an inhomogeneous atomistic charge density $n(R)$ by defining a local $r_s(R)$ from it. However, while such an empirical averaging could result in a very reasonable (due to a sum rule) effective excitation energy in the denominator of the Born-Bethe logarithm, an essential part of excitation-physics behind a sign-dependent term is improperly treated. For close impact ($R$ is small)
the negative swift projectile moves inside atomic orbits. There, the consideration of shift in discrete electron-binding results in a sign-effect {\it opposite} \cite{Kabachnik90,Nagy19} to the above-discussed one. Thus, the cancellation of opposite-sign trajectory-dependent terms may strongly influence the formal enhancement based on an empirical weighting of homogeneous results, considering random collisional conditions. Moreover, due to the mean spacing between lattice-ions in solids, 
application of large-distance dipole fields to projectile--atomic-electron excitation have a limited validity.

Close impact, i.e., an energy- and impact parameter-dependent \cite{Nagy19} small closest 
approach in nuclear collision, indeed plays a decisive role in stopping 
of negative projectiles according to the insightful analysis \cite{Fermi47} of Fermi and Teller which was motivated by the slowing down of negative mesotrons (i.e., muons) in matter (metal, insulator, and gas). Recent {\it ab initio} work for proton and antiproton stopping in $LiF$ target also signals \cite{Bruneval21} the possibility of an {\it opposite} sign-effect, by quantifying the dominant role of small close impact via 
antiproton-made-destabilization of a $p$-type anion orbit in $F^{-}$. The unified treatment of interrelated nuclear collision and electronic excitation is highly desirable, as \cite{Nagy19,Correa12} indicate.

In the light of the above discussion on the sign-dependent term in stopping power, finally we return 
oncemore to the motivating experimental work in \cite{Golovchenko82}. There, besides stating the half-like magnitude
of the observable, it was pointed out that there could be an efficient cancellation between the 
sign-dependent $Z_1 L_1$ term and the next-order negative  ($Z_1^2 L_2$) so-called Bloch term for swift positive projectiles used in experiment. This term is about 
\begin{equation}
Z_1^2\, L_2 \simeq{-\left(\frac{Z_1 v_0}{v}\right)^2}=-\left(\frac{Z_1}{a_0 k}\right)^2 \nonumber
\end{equation}
by considering \cite{Lindhard76} the dominating [$(l=0)\rightarrow{(l=1)}$] angular-momentum
selection with a bare Coulomb force. The dominance of such a contribution in a mathematically convergent series [actually, a sum for Riemann $\zeta(3)$] is compatible 
with the physical form of interaction at short range.
Obviously, due to the $(1/v^2)$-dependence of $L_1$ in our 
Eq.(11),  we arrive at a simple, kinematical \cite{Nagy20}, explanation to terms-cancellation in the positive-ion case.

%\newpage

\section{Summary and outlook}
The nonperturbative kinetic framework for the stopping power of an homogeneous electron gas for 
swift projectiles is implemented. The pair-interaction potential energy required in this framework 
to two-body elastic scattering is based on the screened interaction energy between system particles.
In such an energetically optimized modeling the swift bare projectile interacts with independent screened constituents of a fixed-density interacting many-body target. 
We contrasted the such-obtained theoretical results with independent experimental data 
obtained for Si target by using swift ions. We found harmony with
the reduced value of the stopping power under channeling condition. Similarly, 
the magnitude of the charge-sign effect and its velocity-scaling are in accord with measurements
for the observable. The origin of cancellation between higher-order terms for positive projectiles
is discussed as well. Based on these above, a coherent modeling to further efforts is found.

Recently, in the very promising field of warm
dense plasma \cite{Zylstra15}, measurement of stopping of energetic [$(v/v_0)\simeq{25}]$ protons,
produced via fusion reaction with $D^3He$ fuel, signals that Eq.(1) would need an essentially larger denominator in the logarithmic factor then the conventional estimation with $\Omega_{pl}$
of valence electrons.  The target (t) is a heated solid berillium metal for which $Z_t=4$ and it has two valence electrons. In addition, at
an estimated electron temperature, $k_BT\simeq{32\, eV}$, enhancement in stopping was observed
in comparison with data obtained for cold target. At that, still moderately high, temperature and for 
$(v/v_0)=25$ for protons, the conventional temperature-independent \cite{Brown05} form in Eq.(1) is supposed to be correct for a homogeneous system. However, as was pointed out in a 
single-parametric ($\bar{I}$) Bethe-style \cite{Zylstra15} analysis of random stopping data based on 
 \begin{equation}
 \frac{dE}{dx}\, =\, \frac{4\pi n_0\, Z_1^2\, e^4}{m v^2}\, \ln\left(\frac{2mv^2}{\bar{I}}\right), \nonumber
\end{equation}
with an effective ($\bar{I}$) target characteristics, the $\bar{I}=\hbar \Omega_{pl}$ identification
is not a proper choice to interpret enhancement in stopping data as a function of the temperature.

We speculate that a temperature-dependent extension of our approach might contribute to a
deeper understanding of experimental predictions. Namely, superimposing the random thermalized motion of target electrons ($e$) in Eq.(8) via a new \cite{Ladanyi92} statistical term in 
\begin{equation}
V(K,T)\, =\, - \, Z_1\, v(K,T)\, =\, -Z_1e^2\,  \frac{4\pi\, (K^2 + {\langle{k_e^2}\rangle}) }{K^4 +K^2{\langle{k_e^2}\rangle} + 4\Lambda^4}, \nonumber
\end{equation}
with a thermal-excess, ${\langle{k_e^2}\rangle}\propto{(k_B T m/\hbar^2)}$, 
could increase $(dE/dx)_{T=0}$.
At small temperature we expect a $T$-linear enhancement 
in our Born expression by growing temperature since
\begin{equation}
\left(\frac{dE}{dx}\right)_T \simeq{n_0\, \frac{4\pi Z_1^2 e^4}{m v^2}}\, 
\left(\frac{1}{2}\ln {\frac{2k^2}{\Lambda^2}} + \frac{1}{8}\frac{\langle{k_e^2}\rangle}{\Lambda^2}
\arctan \frac{2k^2}{\Lambda^2}\right),
 \nonumber
\end{equation}
perturbatively. At the experimental conditions (see, above) the second term is roughly 
$1/5$ of the first one. This is not in contradiction with subtracted data \cite{Zylstra15}.
A more quantitative analysis is left to a dedicated paper. 
For a berillium target ($Z_t=4$, with two valence and two $1s$-type electrons in cold metals), the number-density [$n_0(T)$] of effective electrons at fixed volume seems to be a 
Saha-ionization-based parameter. Thus it could be, for berillium, $n_0(T)=2\, n_0$ already at moderate
temperatures due to such \cite{Landau58,Peierls79} ionization.

More generally, and
according to \cite{Zylstra15,Frenje19}, an accurate theory of charged-particle stopping in electron plasmas is a fundamental challenge. It has an impact on our understanding of energy-deposition processes 
and thus on a reasonable estimation for the ignition threshold in fusion research. However, at high temperatures the target ionic component, not considered in the above outlook, may also contribute to the ion-stopping \cite{Frenje19} via nuclear collisions.

\begin{acknowledgments}
We are thankful to Professor P. M. Echenique for useful discussions.
This work was supported partly by the  Grant  PID2019-105488GB-I00 funded by MCIN/AEI/10.13039/501100011033.
\end{acknowledgments}


\newpage

\begin{thebibliography}{00}
%
\bibitem{Golovchenko82}
J. A. Golovchenko, D. E. Cox, and A. N. Goland, Phys. Rev. B {\bf 26}, 2335 (1982).
%
\bibitem{Lindhard54}
J. Lindhard, K. Dan. Vidensk. Selsk. Mat. Fys. Medd. {\bf 28}, No. 8 (1954).
%
\bibitem{Tonks29}
L. Tonks and I. Langmuir, Phys. Rev. {\bf 33}, 195 (1929).
%
\bibitem{Fetter71}
A. L. Fetter and J. D. Walecka, {\it Quantum Theory of Many-Particle Systems} (McGraw-Hill,
London, 1971).
%
\bibitem{Pines61}
D. Pines, {\it The Many-Body Problem} (Benjamin, New York, 1961).
%
\bibitem{Feynman72}
R. P. Feynman, {\it Statistical Mechanics} (Addison Wesley, New York, 1972).
%
\bibitem{Brown05}
L. S. Brown, D. L. Preston, and R. L. Singleton Jr., Phys. Rep. {\bf 410}, 237 (2005).
%
\bibitem{Bonderup81}
E. Bonderup, {\it Lecture Notes on Penetration of Charged Particles through Matter},
2nd ed. (University of Aarhus, Aarhus, 1981).
%
\bibitem{Pines66}
D. Pines and P. Nozieres, {\it The Theory of Quantum Liquids} (Benjamin, New York, 1966).
%
\bibitem{Nagy93}
I. Nagy, A. Arnau, and P. M. Echenique, Phys. Rev. B {\bf 48}, 5650 (1993).
%
\bibitem{Komarov81}
M. A. Kumakhov and F. F. Komarov, {\it Energy Loss and Ion Ranges in Solids}
(Gordon and Breach, New York, 1981).
%
\bibitem{Andersen89}
L. H. Andersen, P. Hvelplund, H. Knudsen, S. P. Moller, J. O. P. Pedersen, U. Uggerhoj,
K. Elsener, and E. Morenzoni, Phys. Rev. Lett. {\bf 62}, 1731 (1989).
%
\bibitem{Sigmund82}
P. Sigmund, Phys. Rev. A {\bf 26}, 2497 (1982).
%
\bibitem{Arista84}
L. de Ferrariis and N. R. Arista, Phys. Rev. A {\bf 29}, 2145 (1984).
%
\bibitem{Bonig89}
L. B\"onig and K. Sch\"onhammer, Phys. Rev. B {\bf 39}, 7413 (1989).
%
\bibitem{Zwicknagel99}
G. Zwicknagel, Ch. Toepffer, and P. G. Reinhard, Phys. Rep. {\bf 309}, 117 (1999).
%
\bibitem{Nagy20}
I. Nagy and I. Aldazabal, Nucl. Instrum. Methods B {\bf 474}, 74 (2020), and references therein.
%
\bibitem{Ziman69}
J. M. Ziman, {\it Elements of Advanced Quantum Theory} (Cambridge University Press, New York, 1969),
pp. 162-164.
%
\bibitem{Mahan81}
G. D. Mahan, {\it Many-Particle Physics} (Plenum Press, New York, 1981).
%
\bibitem{Bonch59}
V. L. Bonch-Bruevich and V. B. Glasko, Sov. Phys. Dokl. {\bf 4}, 147 (1959).
%
\bibitem{Bergara99}
I. Nagy and A. Bergara, J. Phys. : Condens. Matter {\bf 11}, 3943 (1999).
%
\bibitem{Ladanyi92}
K. Lad\'anyi, I. Nagy, and B. Apagyi, Phys. Rev. A {\bf 45}, 2989 (1992), {\bf 46}, 1704(E) (1992).
%
\bibitem{Azevedo01}
G. de M. Azevedo, P. L. Grande, M. Behar, J. F. Dias, and G. Schiwietz, Phys. Rev. Lett.
{\bf 86}, 1482 (2001).
%
\bibitem{Sigmund01}
P. Sigmund and A. Schinner, Phys. Rev. Lett. {\bf 86}, 1486 (2001).
%
\bibitem{Flugge71}
S. Fl\"ugge, {\it Practical Quantum Mechanics I} (Springer-Verlag, Berlin, 1971), Problem 123.
%
\bibitem{Lindhard76}
J. Lindhard, Nucl. Instrum. Methods {\bf 132}, 1 (1976).
%
\bibitem{Quinteros91}
T. B. Quinteros and J. F. Reading, Nucl. Instrum. Methods B {\bf 53}, 363 (1991).
%
\bibitem{Esbensen90}
H. Esbensen and P. Sigmund, Ann. Phys. {\bf 201}, 152 (1990).
%
\bibitem{Pitarke95}
J. M. Pitarke, R. H. Ritchie, and P. M. Echenique, Phys. Rev. B {\bf 52}, 13883 (1995).
%
\bibitem{Kabachnik90}
L. L. Balashova, N. M. Kabachnik, and V. N. Kondratev, Phys. Stat. Solidi (b) {\bf 161}, 113 (1990).
%
\bibitem{Nagy19}
I. Nagy and I. Aldazabal, Adv. Quantum Chem. {\bf 80}, 23 (2019).
%
\bibitem{Fermi47}
E. Fermi and E. Teller, Phys. Rev. {\bf 72}, 399 (1947).
%
\bibitem{Bruneval21}
X. Qi, F.  Bruneval, and I. Maliyov, {\it submitted for publication} (2021).
%
\bibitem{Correa12}
A. A. Correa, J. Kohanoff, E. Artacho, D. S\'anchez-Portal, and A. Caro, Phys. Rev. Lett.
{\bf 108}, 213201 (2012).
%
\bibitem{Zylstra15}
A. B. Zylstra, J. A. Frenje, P. E. Grabowski, C. K. Li, G. W. Collins, P. Fitzsimmons, S. Glenzer,
F. Graziani, S. B. Hansen, S. X. Hu, M. Gatu Johnson, P. Keitner, H. Reynolds, J. R. Rygg,
F. H. S\'eguin, and R. D. Petrasso, Phys. Rev. Lett. {\bf 114}, 215002 (2015).
%
\bibitem{Landau58}
L. D. Landau and E. M. Lifshitz, {\it Statistical Physics} (Pergamon Press, London, !958).
%
\bibitem{Peierls79}
R. Peierls, {\it Surprises in Theoretical Physics} (Princeton University Press, Princeton, 1979).
%
\bibitem{Frenje19}
J. A. Frenje, R. Florido, R. Mancini, T. Nagayama, P. E. Grabowski, H. Rinderknecht, H. Sio,
A. Zylstra, M. Gatu Johnson, C. K. Li, F. H. Seguin, R. D. Petrasso, V. Yu. Glebov, and S. P. Regan,
Phys. Rev. Lett. {\bf 122}, 015002 (2019).
%


\end{thebibliography}
\end{document}